\documentclass[12pt]{iopartA}
\usepackage{epsfig,float}
\usepackage{graphicx}
\usepackage{iopams}

\usepackage{latexsym}

\usepackage[dvips,a4paper]{hyperref}

\makeatletter

\newcommand{\EQ}{\begin{equation}}
\newcommand{\EN}{\end{equation}}
\newcommand{\bea}{\begin{eqnarray}}
\newcommand{\eea}{\end{eqnarray}}

\newcommand{\beq}{\begin{equation}}
\newcommand{\eeq}{\end{equation}}

\newcommand{\intd}{\mathrm{d}}

\newcommand{\ZZ}{\mathbb{Z}}

\newcommand{\ket}[1]{|#1 \rangle}

\newcommand{\brakettt}[3]{\langle #1 | #2 |#3 \rangle}

\makeatother

\begin{document}

\begin{flushright}
SISSA 55/2009/EP
\end{flushright}
\vspace{0.5cm}

\title{Particle spectrum of the 3-state Potts field theory: a numerical study}

\author{L. Lepori$^{1,2}$, G. Z. T\'oth$^{1,2,3}$, G. Delfino$^{1,2}$ }

\address{$^1$ International School for Advanced Studies (SISSA),
Via Beirut 2-4, 34014 Trieste, Italy}
\address{$^2$ INFN Sezione di Trieste}
\address{$^3$ Research Institute for Particle and Nuclear Physics,
Hungarian Academy of Sciences, Pf.\ 49, 1525 Budapest,
Hungary}

\eads{\mailto{delfino@sissa.it}, \mailto{lepori@sissa.it}, \mailto{toth@sissa.it} and \mailto{tgzs@cs.elte.hu}}

\begin{abstract}
The three-state Potts field theory in two dimensions with thermal and magnetic perturbations provides the simplest model of confinement allowing for both mesons and baryons, as well as for an extended phase with deconfined quarks. We study numerically the evolution of the mass spectrum of this model over its whole parameter range, obtaining a pattern of confinement, particle decay and phase transitions which confirms recent predictions. 
\end{abstract}

\maketitle

\section{Introduction}
\label{sec.1}

The $q$-state Potts field theory describes the universality class associated with the spontaneous breaking of the permutation symmetry of $q$ colors. In two spacetime dimensions the theory is defined for $q\le 4$. The transition between the disordered (symmetric) phase and the ordered (broken symmetry) phase occurs as the temperature is varied.
If an external magnetic field is allowed to be switched on, then the theory provides a model of confinement with several interesting features \cite{DeGr}. 
Below critical temperature and in absence of the field, the particle spectrum contains kinks, which interpolate between the $q$ degenerate vacua. The magnetic field breaks (at least partially) the degeneracy of the vacua and those kinks which start or end on vacua that become false get removed from the spectrum. Those multikink configurations which consist of such kinks but start and end on vacua that remain true can give rise to new particles, which will be bound states of the confined kinks.  The role of the kinks is analogous to that of quarks  in chromodynamics, and the kink bound states arising play a role analogous to that of mesons and baryons, therefore the kinks and the bound states are sometimes referred to as quarks, mesons and baryons. 

The mechanism of confinement through the breaking of the degeneracy of discrete vacua is quite general in two dimensions. The Ising case (corresponding to $q=2$) provides the simplest example and was first studied in \cite{McCoyWu} (see \cite{report} for more references). In this as in other cases \cite{msg,LMT,MT}, the single-component order parameter leads to bound states made up of a kink and an anti-kink, i.e.\ only mesonic particles are present. The $q=3$ Potts model provides the simplest example allowing also for baryonic particles (made up of three kinks), as well as for an extended phase in the parameter space in which some kinks are deconfined. Moreover, in this case the renormalization group trajectories originate from a non-trivial fixed point.

A qualitative characterization of the evolution of the mass spectrum in the three- and four-state Potts models for generic values of the temperature and of a magnetic field chosen to act on a single color was given in \cite{DeGr}.
Our aim in the present paper is to verify that picture in the $q=3$ case by numerical calculations. We use the method called the truncated conformal space approach (TCSA) \cite{YZ}, which is suitable for studying mass spectra. Its application requires that the theory is formulated as a perturbation of a conformal field theory. In the $q=3$ case this is the $D_4$ minimal conformal model, and the fact that this is a `non-diagonal' minimal model makes the situation technically somewhat more complicated  compared to the usual applications of the TCSA. Interestingly, our investigation of the off-critical region allows us to fix also some previously unknown data concerning the conformal point.

The paper is organized as follows. In section~\ref{sec.ftd} we introduce the model and discuss the implementation of the TCSA, before presenting the results of the numerical study in section~\ref{sec.ev}. Section~\ref{sec.wmf} is devoted to the comparison with analytic results that can be obtained in the case of weak magnetic field, while section~\ref{sec.c} contains a few final remarks. The results concerning the conformal point are given in the appendix.

\section{Model and numerical method}
\label{sec.ftd}

The three-state Potts model on the lattice \cite{Potts,Wu} is a generalization of the Ising model in which each site variable $s(x)$ 
can take three different values (colors) $1,2,3$. At temperature $T$ and in the presence of an external magnetic field $H$ acting only on the sites with one specific color, say $3$,
the reduced 
Hamiltonian reads as 
\beq
\mathcal{H}=-\frac{1}{T}\sum_{(x,y)}\delta_{s(x),s(y)}-H\sum_{x}\delta_{s(x),3}\ ,
\label{lattice}
\eeq
where the first sum is over nearest neighbours.
At zero magnetic field this Hamiltonian is invariant under the group $S_3$ of permutations of the three colors. If the magnetic field is nonzero, then $\mathcal{H}$ is invariant only under the group $S_2=\ZZ_2$ of permutations of the first two colors.

The system described by the model has two phases, an ordered one and a disordered one. While the disordered phase is characterized by a unique ground state which is invariant under the $\ZZ_2$ symmetry, in the ordered phase the system 
has two possible ground states, which are interchanged by the $\ZZ_2$ transformation. The phase boundary separating the two phases in the $(T,H)$ plane has a first-order and a second-order part. The second-order part runs through the half-plane with negative magnetic field and relates the $q=3$ Potts critical point at $H=0$ to the $q=2$ (Ising) critical point at $H=-\infty$, where the third color is unaccessible. The first-order transition takes place below the critical temperature $T_c$ at zero magnetic field, where three ground states, which are permuted by the $S_3$ symmetry, coexist; in the first-order transition the system also exhibits spontaneous magnetization (the variables $\sigma_{\alpha}(x)= \delta_{s(x),\alpha}-\frac{1}{3}$, $\alpha=1,2,3$ have non-zero expectation value).

The field theoretical description of the scaling limit of the model (\ref{lattice}) begins with the observation that the $S_3$-invariant critical point at $(T,H)=(T_c,0)$, where the first- and second-order transitions meet, belongs to the universality class described by the $D_4$ minimal model of two-dimensional conformal field theory \cite{BPZ,DF,Cardy}. The corresponding value $c=4/5$ of the central charge is the lowest one for which two different modular invariant partition functions can be obtained on a torus  with periodic boundary condition \cite{Cardy}. The first realization, known as $A_5$, contains, with multiplicity one, all the primary operators $\phi$ whose conformal weight $\Delta_\phi$ appears in the Kac table, table~\ref{kac}, and describes the tetracritical point of the Ising model. The second realization, known as $D_4$, contains only a subset of the primaries in table~\ref{kac}, some of them with multiplicity $2$, as expected for the 3-state Potts model, which has a two-component order parameter ($\sum_{\alpha=1}^3\sigma_\alpha=0$). The primaries entering the description of the 3-state Potts model are listed in table~\ref{tab.d4} together with their transformation properties under the group $S_3$, which is the semidirect product of $\ZZ_2$ and $\ZZ_3$; contrary to the situation for the $A_5$ case, some operators in table~\ref{tab.d4} have a non-zero spin $s=\Delta-\bar{\Delta}$.

\begin{table}[t]
\caption{\label{kac} Kac table for $c=4/5$.
}
\begin{indented}
\item
\vspace{2mm}
\begin{tabular}{@{}|c|c|c|c|c|}
\hline
$0$ & $1/8$ & $2/3$ & $13/8$ & $3$  \\
\hline
$2/5$ & $1/40$ & $1/15$ & $21/40$ & $7/5$\\
\hline
$7/5$ & $21/40$  & $1/15$ & $1/40$  & $2/5$   \\
\hline
$3$  & $13/8$ & $2/3$ & $1/8$ & $0$ \\
\hline
\end{tabular}
\end{indented}
\end{table}

\begin{table}[t]
\caption{\label{tab.d4} Primary fields of $D_4$ and their transformation properties.
}
\begin{indented}
\item
\vspace{2mm}
\begin{tabular}{@{}|l|c|c|c|}
\hline
$\phi$ & $(\Delta_\phi,\bar{\Delta}_\phi)$  & $\ZZ_3$ & $\ZZ_2$  \\
\hline
$I$ & $(0,0)$  & $1$ & $1$\\
$\epsilon$ & $(2/5,2/5)$   & $1$  & $1$   \\
$X$  & $(7/5,7/5)$ & $1$ & $1$ \\
$Y$ & $(3,3)$  & $1$ & $1$ \\
$\sigma_1$ & $(1/15,1/15)$  & $\exp( 2\pi\rmi/3)$   &  $\sigma_2$ \\
$\sigma_2$ & $(1/15,1/15)$ & $\exp( 4\pi\rmi/3)$ & $\sigma_1$ \\
$\Omega_1$ & $(2/3,2/3)$ & $\exp( 2\pi\rmi/3)$ & $\Omega_2$ \\
$\Omega_2$ & $(2/3,2/3)$  & $\exp( 4\pi\rmi/3)$ & $\Omega_1$ \\
$J$ & $(7/5,2/5)$  & $1$ & $-1$\\
$\bar{J}$ & $(2/5,7/5)$  & $1$ & $-1$\\
$W$ & $(3,0)$  & $1$ & $-1$\\
$\bar{W}$ & $(0,3)$  & $1$ & $-1$\\
\hline
\end{tabular}
\end{indented}
\end{table}

The 3-state Potts field theory describing the scaling limit of (\ref{lattice}) is obtained by perturbing the $D_4$ conformal theory with the leading thermal operator $\epsilon$ and the leading $\ZZ_2$-even magnetic operator $\sigma_+=(\sigma_1+\sigma_2)/\sqrt{2}$. This leads to the Euclidean action
\EQ
{\cal A}={\cal A}_{D_4}+\tau\int \intd^2x\,\epsilon(x)-h\int \intd^2x\,\sigma_+(x)\,,
\label{action}
\EN
where ${\cal A}_{D_4}$ is the conformal part, $\tau\sim\mbox{mass}^{2-2\Delta_\epsilon}$ is related to the deviation of the temperature from its critical value and $h\sim\mbox{mass}^{2-2\Delta_{\sigma_+}}$ is proportional to the magnetic field. The field theory (\ref{action}) describes a family of renormalization group trajectories flowing out of the origin in the $(\tau,h)$ plane. They are conveniently parameterized by the dimensionless combinations
\beq
\eta_{\pm}=\frac{\tau}{(\pm h)^{(2-2\Delta_\epsilon)/(2-2\Delta_{\sigma_+})}}=
\frac{\tau}{(\pm h)^{9/14}}\,,
\eeq
where the positive (negative) sign applies when $h$ is positive (negative).

As a (1+1)-dimensional theory, (\ref{action}) describes relativistic particles whose mass spectrum, measured in units of the lightest mass, is a function of $\eta_\pm$ only. In order to study numerically the evolution of this spectrum we resort to the method known as truncated conformal space approach (TCSA) \cite{YZ}. This involves first of all considering the theory on a cylinder
of circumference $R$, in such a way that the Hamiltonian operator takes the 
form
\EQ
\label{eq.h}
H=\frac{2\pi}{R}\left(L_0+\bar{L}_0-\frac{c}{12}\right)+
\tau\int_0^R \epsilon(x,0)\,\intd x-h\int_0^R \sigma_+(x,0)\,\intd x\ ,
\EN
where the conformal part is expressed in terms of the zero index generators $L_0$ and $\bar{L}_0$ of the chiral Virasoro algebras with central charge $c=4/5$. 
At the conformal point the eigenvalues of $L_0+\bar{L}_0$ are simply the scaling dimensions $\Delta+\bar{\Delta}$ of the operators, so the conformal space of states coincides with the operator space. Also the matrix elements of the perturbing operators $\epsilon$ and $\sigma_+$ on this space of states can be computed exactly, since they are related to the structure constants $C_{ijk}$ of the conformal operator product expansion as 
\EQ
\brakettt{\phi_i}{\phi_j(0,0)}{\phi_k}=(2\pi/R)^{2\Delta_{\phi_j}}\,C_{ijk}\,.
\label{sc}
\EN
In this way the eigenvalues of (\ref{eq.h}) can be determined by numerical diagonalization of the Hamiltonian matrix on a finite dimensional subspace of the conformal space. Obviously, the truncation of the space of states puts upper bounds on the number of levels and on the values of $R$ which are numerically accessible. Nevertheless, with a number of states which is very reasonable from the point of view of computer time, it is normally possible to obtain a sufficiently accurate description of many energy levels in the lower part of the spectrum, up to values of $R$ for which the asymptotic regime relevant for the theory on the plane is already visible. 

After its appearance \cite{YZ}, the TCSA has been widely used, including for the study of multiple perturbations \cite{DMS}, non-minimal models \cite{FRT} and theories with boundaries \cite{DPTW} (see e.g. \cite{thesis} for additional references). Its use for the case at hand would not present special difficulties were not for the following point. Although minimal (i.e. containing a finite number of operator families), the conformal theory we start from does not belong to the series of minimal models (known as $A$-series) for which the structure constants appearing in (\ref{sc}) are known completely from the work of Dotsenko and Fateev \cite{Dotsenko}. The three-state Potts conformal point, as already said, belongs to the $D$-series of minimal models, and for this series not all the structure constants are completely known; more precisely, some of them are known only up to signs \cite{scalarconst, ocneanu}. Of course, these signs are themselves essential for the determination of the energy spectra through the TCSA. As a matter of fact, we managed to determine them (at least those needed for our purposes) requiring that the output of the TCSA is physically meaningful. The results are given in the appendix.

\section{Evolution of the particle spectrum}
\label{sec.ev}

\begin{figure}
\begin{indented}
\item
\includegraphics[
  scale=0.4]{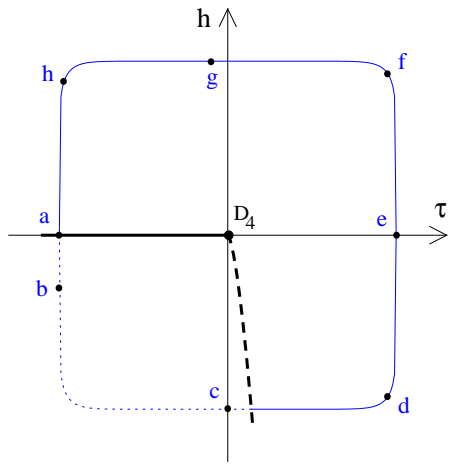}
\end{indented}
\caption{\label{sqr}
{\em Phase diagram of the field theory (\ref{action}). The thick lines 
correspond to the first-order (continuous line) and second-order (dashed line) 
phase transitions. Dots on the closed path $\tau^{14} + |h|^9 =\mbox{constant}$
mark the points corresponding to the spectra shown in figures \ref{fig.ordered}a--h.
The ordered phase corresponds to the dotted part of the path.}
}
\end{figure}

We used the TCSA to follow the evolution of the lower part of the spectrum of the Hamiltonian operator (\ref{eq.h}) in the whole range of the parameters $\eta_\pm$. More precisely, we numerically determined the spectrum at 80 points along the curve $\tau^{14}+|h|^9=\mbox{constant}$ shown in figure~\ref{sqr}: each renormalization group trajectory flowing out of the origin in the $(\tau,h)$ plane and corresponding to a specific value of $\eta_+$ or $\eta_-$ intercepts such a closed curve at a specific point, in such a way that making a round trip along the curve amounts to spanning all the trajectories. Since {\em mass ratios} are fixed along a given trajectory, it is sufficient to determine them at a single point.

For the study of the mass spectrum it is sufficient to consider the subspace of states with vanishing total momentum on the cylinder, namely the operators with zero spin\footnote[1]{Notice that also the primaries with non-zero spin in table~\ref{tab.d4} contribute scalar descendants.}. Moreover, one can treat the parts of the Hamiltonian which are even and odd under the $\ZZ_2$ symmetry separately. We included in our numerical calculations conformal states up to level 8 of descendance within the conformal family associated to each primary operator. This corresponds to $2426$ states in the even sector and $1829$ states in the odd sector. 

Examples of the results we obtained for the spectrum at different values of $\eta_\pm$ are given in figures~\ref{fig.ordered}a--h. They show, as a function of the cylinder circumference $R$, the energy differences $E_i-E_0$, where $E_i$ is the energy of the $i$-th level, with the ground state corresponding to $i=0$. As a consequence, in these figures the ground state coincides with the horizontal axis. Due to the absence of phase transitions on the cylinder, the ground state is always unique at finite $R$. When other levels are seen to approach the horizontal axis as $R$ increases, the spectrum refers to values of $\tau$ and $h$ corresponding to the ordered phase (with degenerate ground states) on the plane ($R=\infty$).

The energy levels behave as $1/R$ in the conformal limit $R\to 0$ (see (\ref{eq.h})) and approach constant values as $R\to\infty$. These asymptotic constants determine the particle spectrum of the theory on the plane, the one we are interested in in this paper. The lowest nonzero asymptotic value determines the mass of the lightest particle (or multi-particle) state with quantum numbers compatible with the boundary conditions chosen on the cylinder geometry. In this paper we always work with periodic boundary conditions, which select states with zero topological charge, and determine the mass of the kinks (which are topologically charged) in the ordered phase from that of the kink-antikink states. 

If $m_1$ is the mass of the lightest particle, the continuous part of the spectrum on the plane starts at $2m_1$. On the cylinder, the continuum breaks into infinitely many discrete levels which become dense when $R\to\infty$: the lowest one (the threshold) corresponds to a pair of particles at rest, the others to a pair with momenta $p$ and $-p$. Of course, in our truncated numerical approach, only a finite number of these `momentum lines' are visible. 
In general, the theory on the plane possesses several stable particles with 
masses $m_i$, and the total spectrum is made up of all the states resulting from the combination of these particles.

With these remarks in mind, it is not difficult to see that the numerical results indeed confirm the evolution of the particle spectrum predicted in \cite{DeGr}. Referring the reader to that paper for the detailed arguments at the origin of the prediction, we recall here the main features and how they appear in the TCSA. 

\begin{figure}
\begin{tabular}{cc}
\includegraphics[
  scale=0.7]{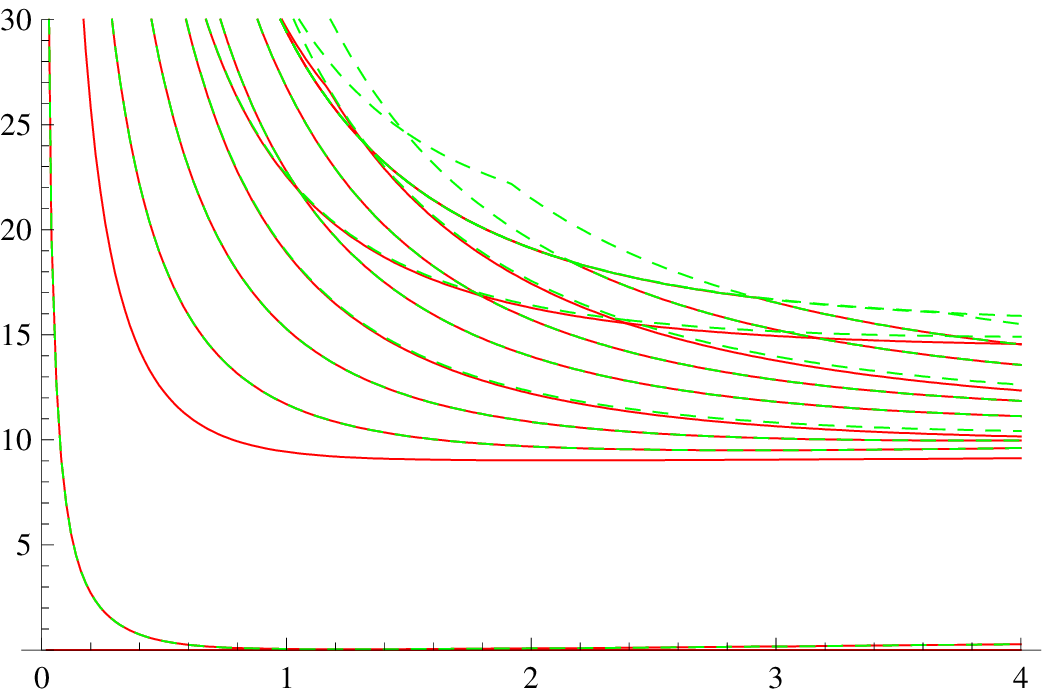} & 
\includegraphics[
  scale=0.7]{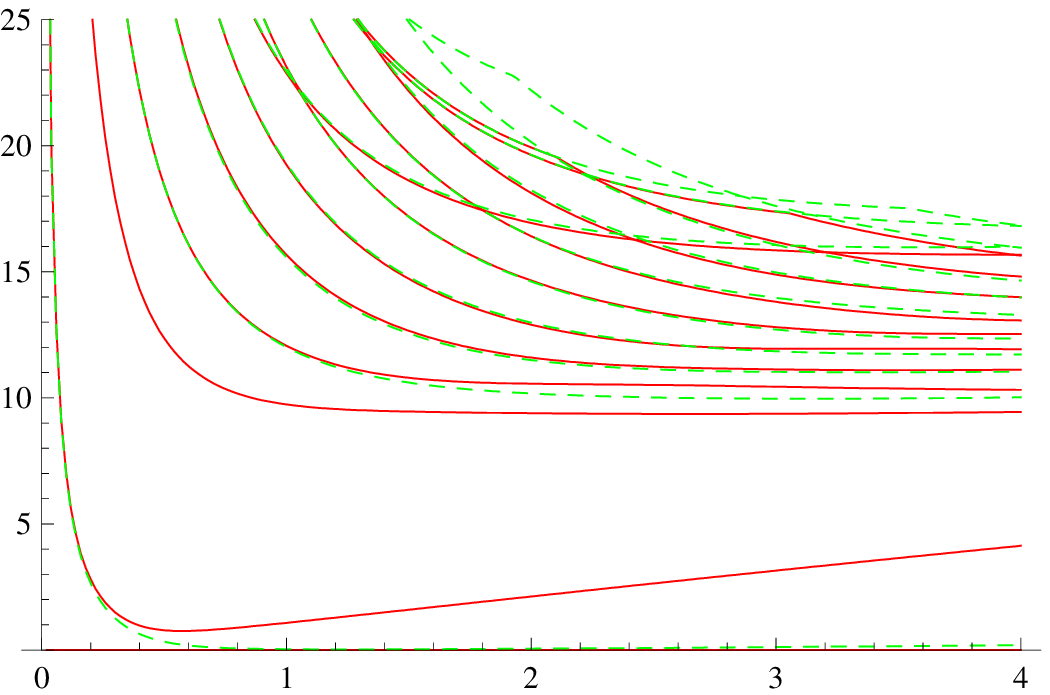}\\
(a) $\eta_\pm=-\infty$ 
 & (b) $\eta_- =-2.168$  
\end{tabular}
\end{figure}

\begin{figure}
\begin{tabular}{cc}
\includegraphics[
  scale=0.7]{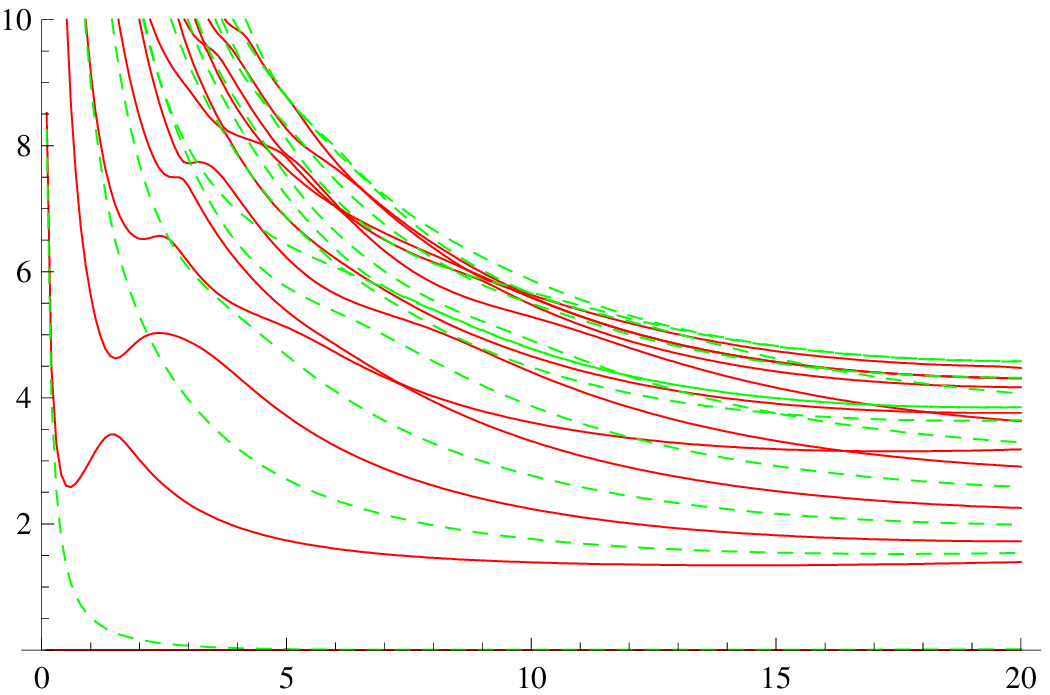} & 
\includegraphics[
  scale=0.7]{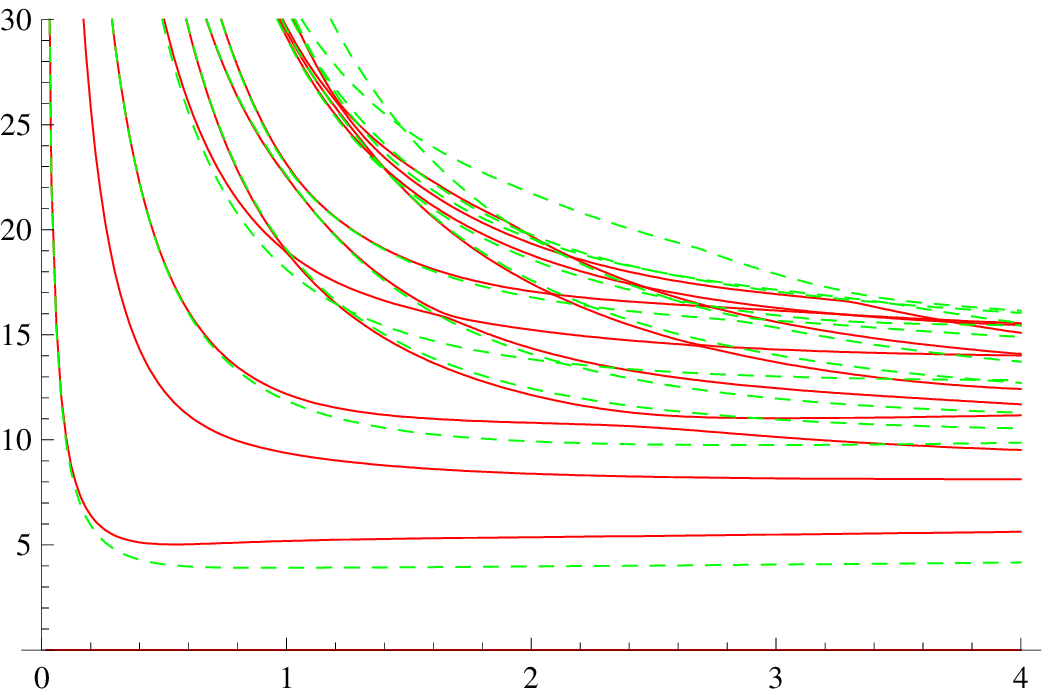}\\
(c) $\eta_-=0$ 
 & (d) $\eta_-=1$ 
\end{tabular}
\end{figure}

\begin{figure}
\begin{tabular}{cc}
\includegraphics[
  scale=0.7]{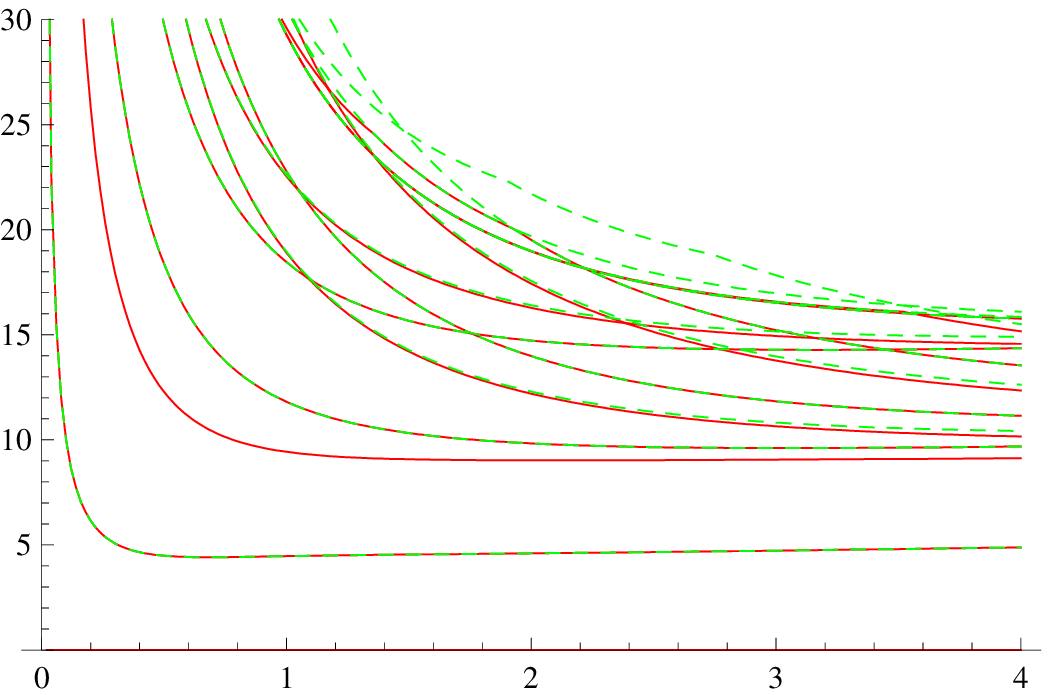} & 
\includegraphics[
  scale=0.7]{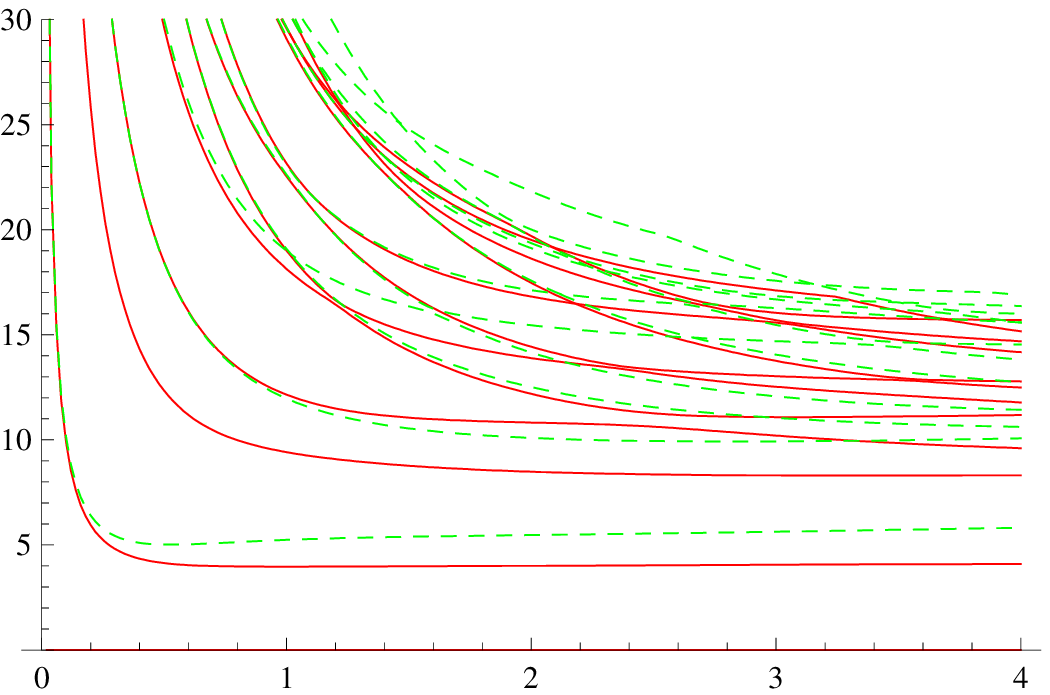}\\
(e) $\eta_\pm=+\infty$ 
& (f) $\eta_+=1 $ 
\end{tabular}
\end{figure}

\begin{figure}
\begin{tabular}{cc}
\includegraphics[
 scale=0.7]{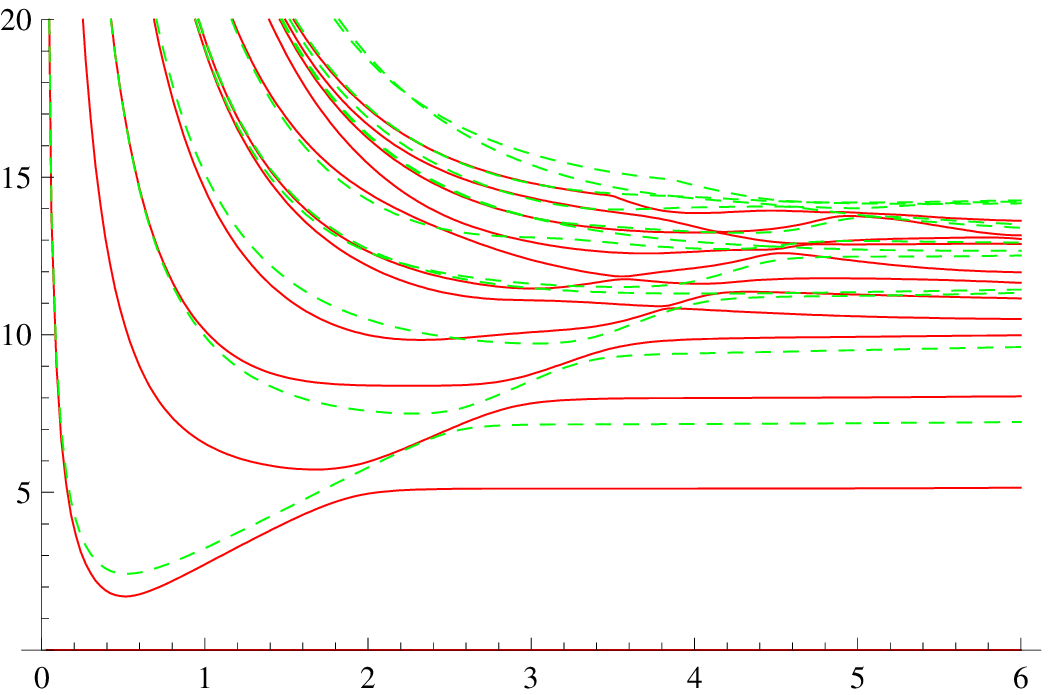} & 
\includegraphics[
  scale=0.7]{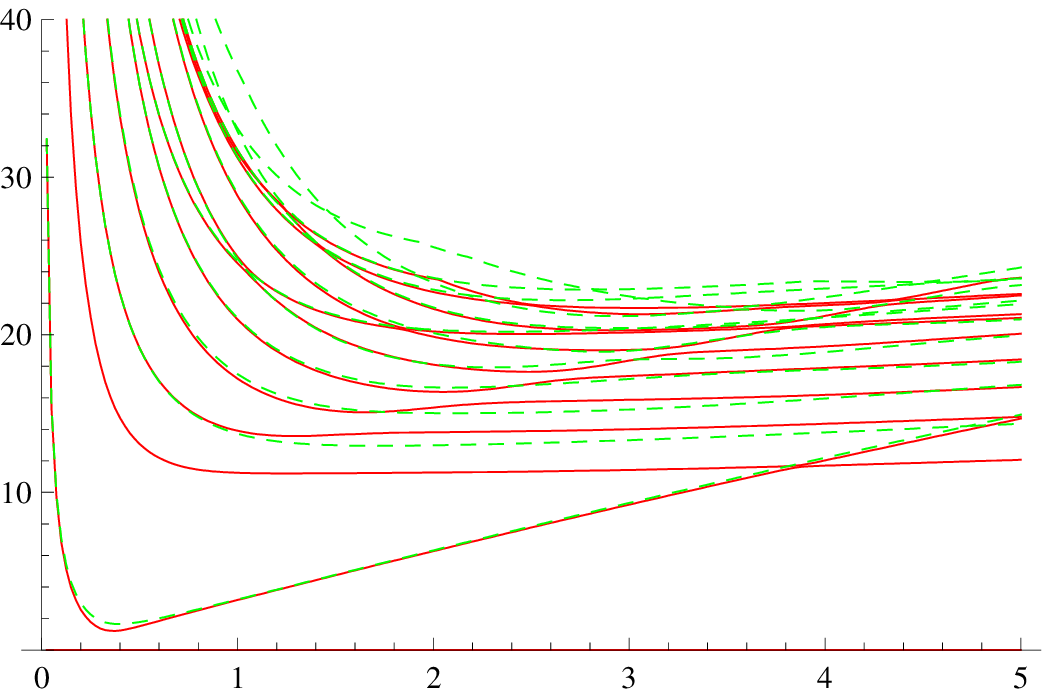}  \\
(g) $\eta_+=-0.1$
& (h) $\eta_+=-1.07$ 
\end{tabular}
\caption{\label{fig.ordered}
{\em The low lying energy differences $E_i-E_0$ as functions of $R$ at the points a--h along the closed path of figure~\ref{sqr}. Even levels are in red; odd levels are in green with dashed lines. Units refer to a fixed mass scale.}
}
\end{figure}

\subsection{The ordered phase}

We start our discussion at low temperature ($\tau<0$) and zero magnetic field, where the model is on the first-order phase transition line. 
Here the system  on the plane exhibits a spontaneous breaking of the $S_3$ symmetry and has three degenerate ground states $\ket{0_\alpha}$, $\alpha=1,2,3$. The spectrum of elementary excitations is known \cite{CZ} to contain only the kinks $K_{\alpha\beta}$, $\alpha\ne\beta$, of equal mass $m$, which interpolate between the ground states $\alpha$ and $\beta$. 
The finite volume spectrum calculated by the TCSA is shown in figure~\ref{fig.ordered}a. Two superimposed levels are clearly seen to approach the horizontal axis to produce the asymptotic triple ground state degeneracy. 
The splitting of the vacua is due to tunneling effects and, up to truncation errors, decreases exponentially at large $R$. The unique finite volume ground state belongs to the $\ZZ_2$-even sector; it becomes $\ket{0_3}$ asymptotically, while the other two levels go into $\ket{0_1}\pm\ket{0_2}$. The single-kink states do not appear due to the periodic boundary conditions, but two-kink states $K_{\alpha\beta}K_{\beta\alpha}$ as well as three-kink states $K_{\alpha\beta}K_{\beta\gamma}K_{\gamma\alpha}$ are clearly visible. 

When, starting from this situation, a magnetic field of the form specified in (\ref{action}) is switched on, the vacuum $\ket{0_3}$ is no longer degenerate with the other two: in particular, for $h<0$ its energy is higher and it becomes a false vacuum. In the finite volume this translates into an energy level linearly increasing with $R$, a feature perfectly  visible in figure~\ref{fig.ordered}b. Since finite energy excitations cannot start from or end on a false vacuum, the kinks 
$K_{13}$, $K_{23}$ and their antikinks get removed from the spectrum, while $K_{12}$ and $K_{21}$ survive. The multi-kink states in which $\gamma=3$ enters only as an internal index preserve a finite energy in the negative field. This is the case, in particular, of the state $K_{13}K_{32}$, which however cannot remain a two-kink excitation at $h<0$, since the false nature of the intermediate vacuum prevents the two kinks from moving arbitrarily far apart. They are instead confined into the topologically charged `mesons' $\pi_{12}^{(n)}$, with $n$ a positive integer indexing mesons of increasing mass. While the mesonic spectrum becomes dense above $2m$ as $h\to 0^-$, the number of stable mesons is expected to decrease as one moves away from the first-order transition line. Indeed, since the mass splittings increase in the process, more and more mesons become unstable crossing the decay threshold $3m$, above which the decay channel $\pi_{12}^{(n)}\to K_{12}K_{21}K_{12}$ opens.

The pattern of the levels visible in figure~\ref{fig.ordered}b is consistent with these expectations. Because of the periodic boundary conditions, the topologically charged particles $K_{12}$ and $\pi_{12}^{(n)}$ and their anti-particles appear only through neutral combinations\footnote{The topologically charged mesons, whose stability is specific of the Potts model with $q=3$, were not mentioned in \cite{DeGr}. They have been recently discussed in \cite{Rutkevich}.}. So the mass gap corresponds to the $K_{12}K_{21}$ threshold and the levels converging at $3m$ to the $\pi_{12}^{(1)}K_{21}$ threshold.
Figure~\ref{fig.ordered}c shows that at $\eta_-=0$ the theory still is in the ordered phase (the single-particle excitations do not appear, so they are kinks). The absence of lines converging at three times the mass of the kinks now indicates that no stable topologically charged meson is left so far away from the first-order transition. The non-monotonic behaviour of the lowest lines in the even sector is a residual manifestation of the false vacuum visible in figure~\ref{fig.ordered}b, which is now very unstable.

\subsection{The second-order phase transition}
\label{sec.p}

The reduction of the mass gap from figure~\ref{fig.ordered}b to figure~\ref{fig.ordered}c is consistent with the expectation that the ordered phase ends at a critical value $\eta_-^c$ where the kink mass vanishes and a second-order transition of Ising type takes place. The value $\eta_-^c$ had been naively set to 0 in \cite{DeGr}. There is, however, no symmetry (like duality in the tricritical Ising model) which selects this value and it was observed in \cite{isingperc} that, since the Potts Curie temperature is larger at $q=2$ than at $q=3$, $\eta_-^c$ is more likely positive. This latter expectation is indeed confirmed by the TCSA, which neatly shows a phase transition at $\eta_-^c\simeq 0.14$ and confirms that it corresponds to a crossover from $q=3$ to $q=2$ criticality.

\begin{table}
\begin{indented}
\item
\begin{tabular}{@{}r|rrrrrrr}
 &   ~~$0$  & ~$1$  & ~$2$  & ~$3$  &  ~$4$ & $5$  & $\cdots$ \\
\hline
$(0,0)$ & $1$  & $0$  & $1$  & $1$  & $4$  &  $4$ & $\cdots$ \\
$(1/2,1/2)$ & $1$  & $1$  & $1$  & $1$  & $4$   & $4$  & $\cdots$ \\
$(1/16,1/16)$ & $1$  & $1$  & $1$  & $4$  & $4$ & $9$  & $\cdots$ \\
\hline
\end{tabular}
\caption{Multiplicities of low lying states of zero spin in the representations of the conformal algebra entering the Ising model; $n$ is displayed in the top row.
\label{tab1}}
\end{indented}
\end{table}

To see this point notice that the signature of a crossover to Ising criticality in the finite volume spectrum is that for sufficiently large values of $R$ the energy levels behave again as $1/R$, or more precisely as
\begin{equation}
\label{eq.A}
E_i-E_0 \sim \frac{2\pi}{R}(\Delta_i+\bar{\Delta}_i),
\end{equation}
where 
$\Delta_i+\bar{\Delta}_i$ are scaling dimensions of Ising operators. Since we are considering the zero momentum sector, and on the cylinder the momentum coincides with the spin, only those states with $\Delta=\bar{\Delta}$ have to be taken into account. The Ising model with periodic boundary conditions contains three representations of the conformal algebra with highest weights $\Delta=0,1/2,1/16$. The scaling dimensions entering (\ref{eq.A}) are then $2(\Delta+n)$, where $n=0,1,2,\dots$ labels the conformal levels. The multiplicities of the states with these dimensions are shown in table \ref{tab1}.

\begin{figure}
\begin{tabular}{cc}
\includegraphics[
  scale=0.32]{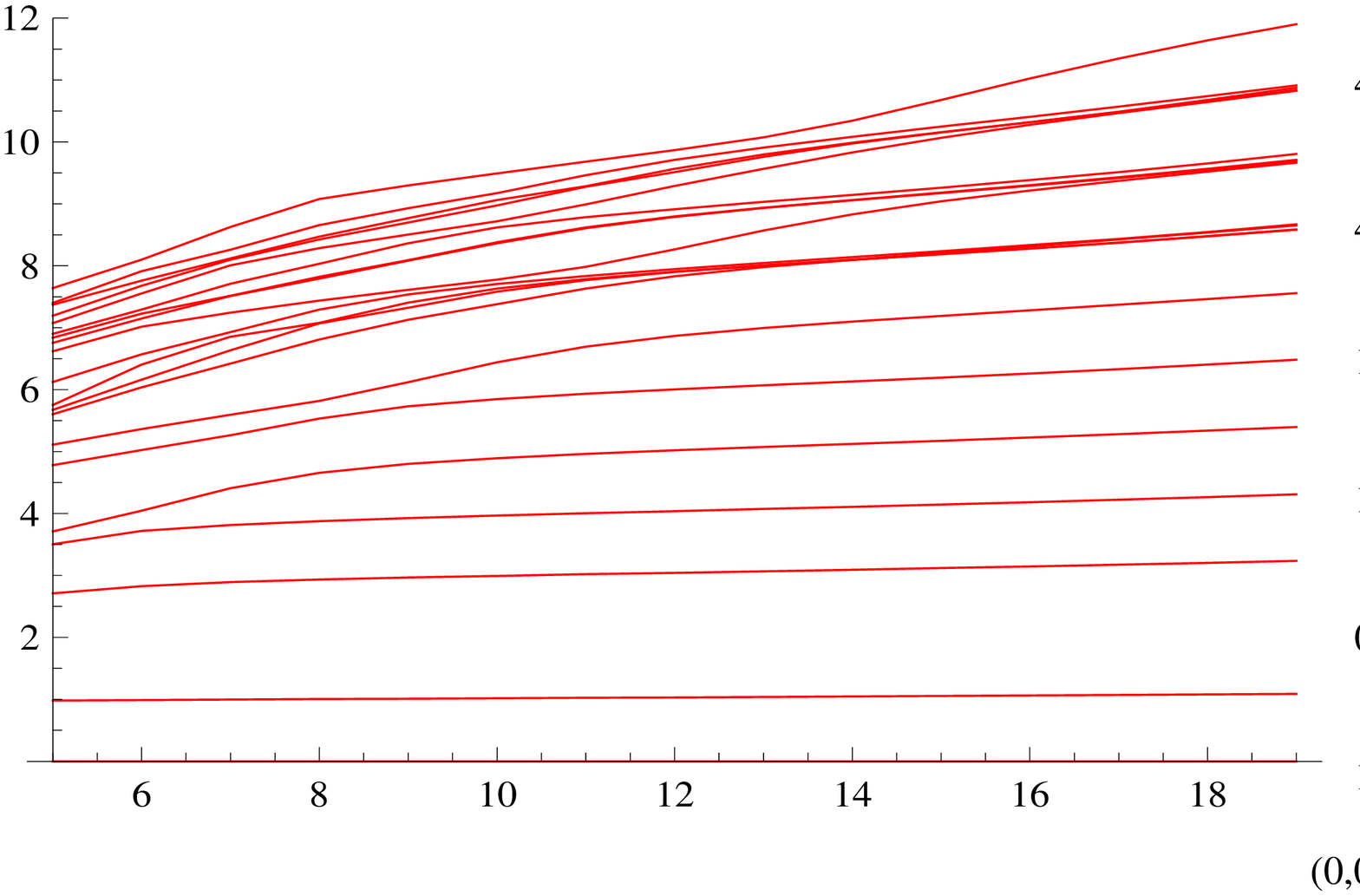} & 
\includegraphics[
  scale=0.32]{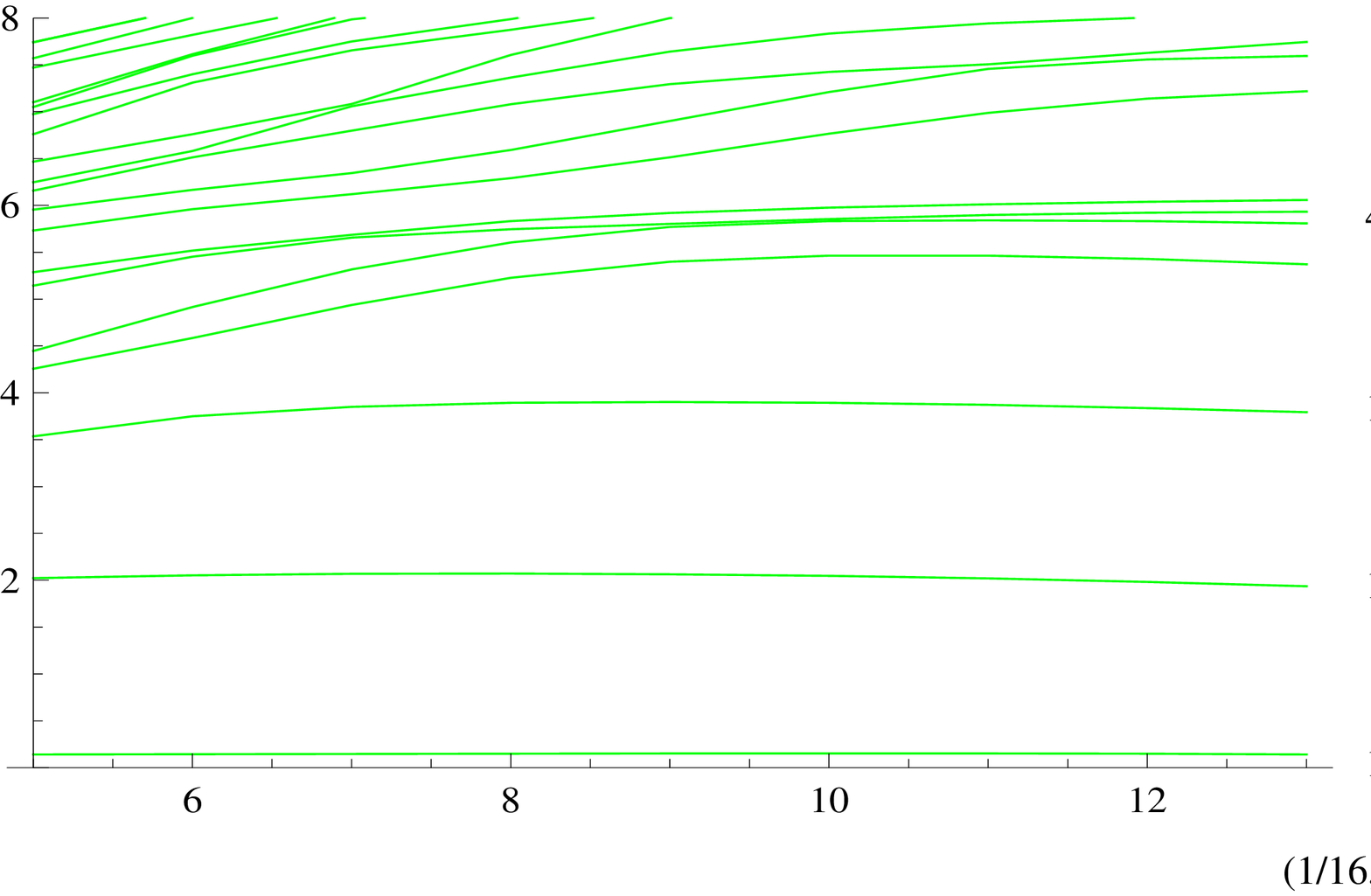}\\
(a) even sector & (b) odd sector
\end{tabular}
\caption{\label{fig.crit}
{\em $\frac{R}{2\pi}(E_i-E_0)$ as functions of $R$ for the lowest values of $i$ in the (a) even and (b) odd sectors at the second-order phase transition point. Degeneracies predicted by the critical Ising model (Table \ref{tab1}) are shown on the right hand side.}
}
\end{figure}

\begin{figure}
\begin{indented}
\item
\begin{tabular}{c}
\includegraphics[
  scale=0.7]{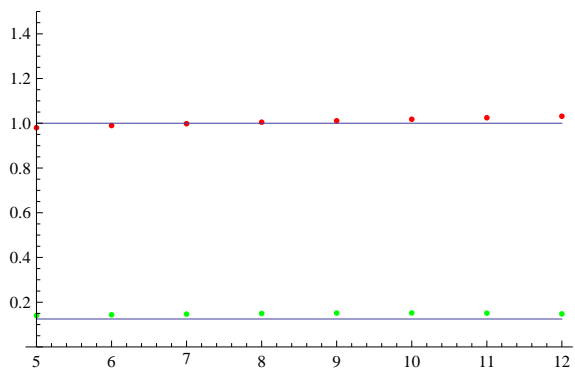} 
\end{tabular}
\end{indented}
\caption{\label{fig.delta}
{\em The lowest energy difference $\frac{R}{2\pi}(E_i-E_0)$ in the two sectors as a function of $R$ at the second-order phase transition point. Data are shown with dots, the constants $2\Delta$ equal $1/8$ and $1$ predicted by the critical Ising model are shown with continuous  lines. }
}
\end{figure}

Figure \ref{fig.crit} shows $\frac{R}{2\pi}(E_i-E_0)$ as functions of $R$ in the even and odd sectors at the point $\eta_-=0.14$ (the same absolute ground state energy is subtracted in both sectors). According to (\ref{eq.A}) these functions should be approximately constant for sufficiently large values of $R$. The data are in good agreement with this expectation and also
the location and the multiplicities of the lines agree well with the data of the critical Ising model. For the lowest energy difference in each sector the accuracy of the data can be appreciated in figure~\ref{fig.delta}.

Similar studies of second-order phase transitions by means of the TCSA were done in \cite{BPTW,Toth}. A second-order phase transition was also observed in the tricritical Ising model \cite{LMT}, however its type (which is known to be Ising) could not be established numerically because of large truncation errors.

\subsection{The disordered phase}

When, starting from the first-order transition line, the magnetic field is switched on with a {\em positive} sign, $\ket{0_3}$ remains as the unique true vacuum of the theory. We thus enter the disordered phase in which all kinks are confined inside topologically neutral mesons originating from $K_{31}K_{13}\pm K_{32}K_{23}$, and baryons originating from $K_{31}K_{12}K_{23}\pm K_{32}K_{21}K_{13}$, where we have taken linear combinations with definite $\ZZ_2$-parity. We keep the notation of \cite{DeGr} and denote $\pi_0^{(n)}$ and $\pi_1^{(n)}$ the even and odd mesons, respectively, and $p_\pm^{(n)}$ the baryons with parity $\pm 1$. In all cases the positive integer $n$ indexes particles with increasing mass, and again the spectra of these particles become dense (above $2m$ for the mesons and above $3m$ for the baryons) in the limit $h\to 0$; as we move away from the first-order transition line more and more particles should cross the two-meson decay thresholds and become unstable.

These expectations are fully consistent with the numerical spectrum shown in figure~\ref{fig.ordered}h, which exhibits a large number of lines corresponding to single-meson states and a doubly degenerate linearly rising line corresponding to the two false vacua. A linearly rising line-like feature with the same slope can also be seen somewhat higher, which is the signature of some topologically neutral configuration that is unstable because it is supported by the false ground states\footnote{See \cite{DGM} for a discussion of (possibly stable) particles above threshold.}. A line corresponding to the baryon $p_+^{(1)}$ is also present, although it is not very clearly seen because of truncation errors and the presence of many other lines in its vicinity. 

At $\eta_+=+\infty$ the theory is integrable and is known \cite{KS,SashaPotts} to possess a doublet of massive particles as the only single-particle excitations. By duality\footnote{See \cite{DeGr} for the relation between the scattering theories at $\tau>0$ and $\tau<0$ in zero field.} their mass coincides with the mass of the kinks at $\eta_\pm=-\infty$. It is expected that the two lightest mesons $\pi_0^{(1)}$ and $\pi_1^{(1)}$ are the only products of kink confinement which do not decay as $\eta_+$ increases, and that they become the above mentioned doublet at $\eta_+=+\infty$. This prediction is confirmed by the numerical calculations. The reduction in the number of stable single-particle states as $\eta_+$ increases can be appreciated in figures~\ref{fig.ordered}g and \ref{fig.ordered}f. In the latter only the two lightest mesons are left below threshold; their mass difference vanish at $\eta_+=+\infty$ (figure~\ref{fig.ordered}e) and changes sign with the magnetic field (figure~\ref{fig.ordered}d).

\subsection{Complete evolution}

\begin{figure}
\begin{center}
\includegraphics[
  scale=0.52]{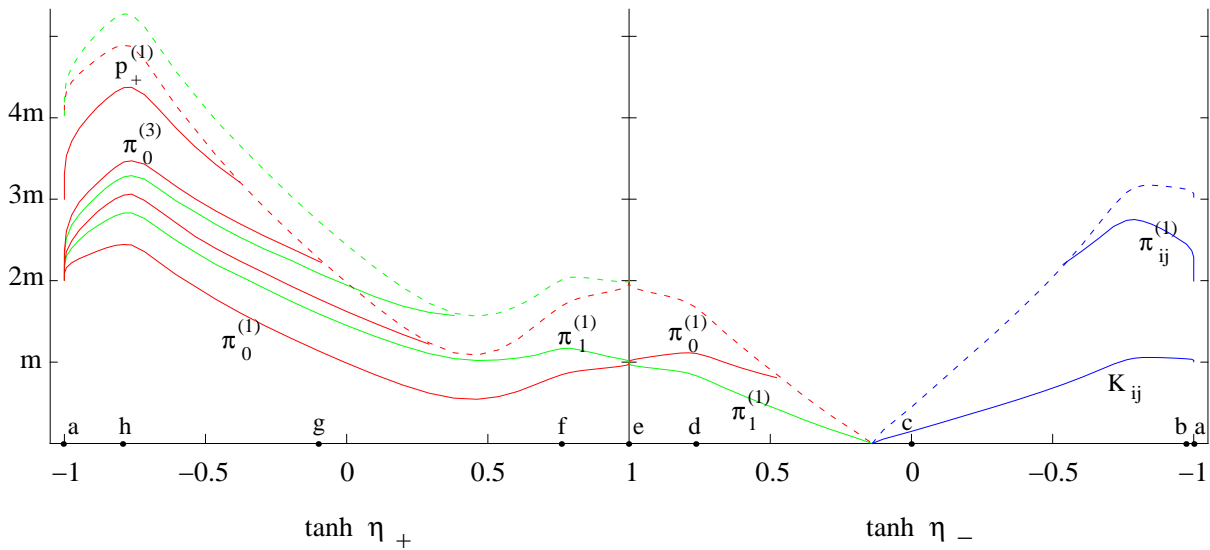}
\end{center}
\caption{\label{mass}
{\em Masses of the lightest mesons $\pi_0^{(1)}$, $\pi_1^{(1)}$, $\pi_0^{(2)}$, $\pi_1^{(2)}$, $\pi_0^{(3)}$, the lightest baryon $p_+^{(1)}$, the elementary kinks $K_{ij}$ and the lightest mesonic kinks $\pi^{(1)}_{ij}$, along the closed path of figure \ref{sqr}. Even particles are shown in red, odd particles in green and the kinks in blue. Dashed lines show the stability thresholds for even particles (twice the mass of the lightest particle), odd  particles (mass of the lightest even particle plus that of the lightest odd particle) and mesonic kinks (mass of $K_{ij}$ plus that of $\pi^{(1)}_{ij}$) in red, green and blue, respectively. The points $a$-$h$, having a correspondence in figures \ref{sqr} and \ref{fig.ordered}, are also marked. }
}
\end{figure}

The systematic collection and analysis of the TCSA spectra at regular small intervals along the closed path of figure~\ref{sqr} allows us to follow the evolution of the lightest particles of the theory over the whole range of the parameters $\eta_\pm$. The result is shown in figure~\ref{mass} and shows a remarkable agreement with the qualitative prediction given in figure~8 of \cite{DeGr}. The data show how the confinement of kinks at $h>0$ produces both mesons and baryons, and how all the mesons, excluding the two lightest, and all the baryons reach the decay thresholds as $\eta_+$ increases. The mesons $\pi_0^{(1)}$ and $\pi_1^{(1)}$ are the only stable particles of the theory in a sufficiently wide range around $\eta_\pm=+\infty$, the  point of enhanced ($S_3$) symmetry where their mass trajectories cross. At $h<0$ the even meson $\pi_0^{(1)}$ is no longer the lightest particle of the theory, and decays before the second-order transition point, where the mass of $\pi_1^{(1)}$ vanishes, is reached. In the ordered phase the spectrum of stable particles contains the deconfined elementary kinks $K_{12}$, $K_{21}$ and, close enough to the first-order transition point $\eta_\pm=-\infty$, the mesonic kinks $\pi^{(n)}_{12}$, $\pi^{(n)}_{21}$. 

Of course, more and more particles emerge from the thresholds and become stable as $\eta_\pm\to-\infty$. Figure~\ref{mass} includes only the five lightest topologically neutral mesons, the lightest baryon and the lightest mesonic kink. The behaviour of the curves at $h=0$ follows from the discussion in the next section.

\section{Weak magnetic field}
\label{sec.wmf}

A number of analytic results can be obtained at $h=0$, where the theory is integrable, and for small $h$, using perturbation theory around the integrable case. In this section we compare some of these results with the numerical ones that we obtain using the TCSA.

At $h=0$ and, by duality, for both signs of $\tau$, the vacuum energy density and the mass of the elementary excitations can be written as 
\EQ
\epsilon_0=a\,\tau^{10/6},\hspace{1cm}m=b\,\tau^{5/6},
\EN
with $a$ and $b$ dimensionless constants whose exact values are known to be\footnote{Throughout the paper we use the normalization of the operators in which $\langle\phi(x)\phi(0)\rangle\to|x|^{-4\Delta_\phi}$ as $|x|\to 0$.} \cite{Fateev}
\EQ
a=-5.856..\,,\hspace{1cm}b=4.504..\,.
\EN
The measure of the slope of $E_0$ and of the mass gap at large $R$ in the TCSA gives $a\approx -5$ and $b\simeq 4.5$.
We are able to determine $a$ only with a low accuracy, a circumstance that is not unexpected: as a general experience absolute energies are given less precisely by the TCSA than mass gaps (energy differences).

Consider now the high temperature phase, $\tau>0$. The first-order correction to $\epsilon_0$ is proportional to the spontaneous magnetization, which is zero.
The second-order correction reads
\beq
\delta\epsilon_0 = -\frac{h^2}{2}\int \intd^2x\langle\sigma_+(x)\sigma_+(0)\rangle
=d\,h^2 \tau^{-13/9}\,.
\eeq
The susceptibility was computed in \cite{DC} (see also \cite{DCGJM}, appendix B) within the two-particle approximation in the form factor approach, with the result $d\simeq-0.135$, which compares very well with the result $d\simeq-0.133$ that we obtain using the TCSA.

Let us denote as
\EQ
\delta m_i^{(1)}=e_i\,h+f_i\,h^2 +..
\EN
the correction to the mass of the meson $\pi_i^{(1)}$, $i=0,1$. At first order one has \cite{DMS,DeGr}
\beq
e_i=-\frac{1}{m}\brakettt{\pi_i^{(1)}(0)}{\sigma_+(0)}{\pi_i^{(1)}(0)}\ .
\eeq
These matrix elements are known (see \cite{DCGJM} and references therein) in a basis $A,\bar{A}$ in which  $\pi_i^{(1)}=(A+(-1)^i\bar{A})/\sqrt{2}$, and read
\EQ
e_{0}=-e_{1}=-0.7036..\,\tau^{-13/18}\,.
\EN
Our TCSA determination $e_{0}=-e_{1}\simeq-0.71..\,\tau^{-13/18}$ is in reasonable agreement with this exact result. As regards the second-order mass corrections, we obtain the numerical results $f_0\approx 0.09\,\tau^{-41/18}$ and $f_1\approx 0.12\,\tau^{-41/18}$. These numbers probably have considerable errors, nevertheless it will be interesting to compare them with analytic results when these are available. 

Turning to the low temperature phase, the confinement of kinks at leading order in the magnetic field can be described within the non-relativistic framework \cite{DeGr}. The detailed analysis has been recently performed in \cite{Rutkevich} for the mesonic spectrum, both at low energy and in the semiclassical limit, as a generalization of the Ising case. As regards the low energy part of the spectrum at $h>0$, the linear confining potential in one dimension produces mesons with masses $m_i^{(n)}$ which deviate from twice the kink mass by a term $c_i^{(n)}h^{2/3}$; the constants $c_i^{(n)}$ are proportional to the zeros of the Airy function for the odd mesons ($i=1$), and to the zeros of the derivative of the Airy function for the even mesons ($i=0$) \cite{Rutkevich}. 

A precise quantitative investigation of this region of the phase diagram in which the mesonic spectrum tends to a continuum requires a numerical accuracy higher than that of our TCSA data. Nevertheless, we found that for moderately large values of the magnetic field ($\eta_+\leq-1)$ the data for the masses of the five lightest mesons are described relatively well by the $h^{2/3}$ power law; the exponent that we could extract by fitting a power function to the data is approximately $0.7$. As regards the prefactors, we obtain very approximately $c_1^{(1)}/c_0^{(1)}\approx 2$, to be compared with the value close to $2.3$ corresponding to the analytic prediction.

\section{Conclusion}
\label{sec.c}

We studied the scaling limit of the two-dimensional three-state Potts model as a function of temperature and magnetic field. This has been done directly in the continuum limit, considering the perturbation of the $D_4$ minimal model of conformal field theory by its leading thermal and magnetic operators. The magnetic perturbation was chosen to be coupled to one color only, in such a way to leave a residual permutation symmetry in the first two colors. We studied the particle spectrum of the theory in the whole plane of the coupling constants using the numerical method called the truncated conformal space approach, with the aim of verifying the qualitative predictions made in \cite{DeGr}. 

Our results confirm these  predictions, clearly showing that kink confinement produces both mesonic and baryonic particles and that deconfined kinks survive within an extended ordered phase whose boundary contains a first-order and a second-order part. In particular, we determined numerically the location of the 
second-order transition, which is not fixed by symmetry. The evolution of the masses of the first few lightest mesonic and baryonic states has been followed through the whole parameter range of the theory, exhibiting the decay pattern and confirming that the two particles of the disordered phase in zero field are the two lightest mesons produced when breaking the ordered phase by a positive field.

The implementation of the numerical method required the knowledge of the structure constants of the thermal and magnetic operators in the $D_4$ conformal field theory. We determined the signs of the structure constants which were left undetermined in \cite{scalarconst, ocneanu}.
The integrability of the model at zero magnetic field makes it possible to obtain analytic results for the particle masses and for the vacuum energy density at zero and small magnetic fields. We compared our numerical results  with such analytic results at high temperature, finding a good agreement. We also presented numerical results for the second-order corrections of the meson masses, which could be useful for comparison with future analytic results. We were able to partially confirm recent results of \cite{Rutkevich} for the meson masses at low temperature 
and small magnetic field. This region of the phase diagram is numerically more demanding because the mesonic spectrum becomes dense as the magnetic field approaches zero, so that a complete verification requires a numerical accuracy larger than that of the present study. Normally this can be achieved by increasing the truncation level.

\section*{Acknowledgments}

The authors thank Paolo Furlan, Giuseppe Mussardo, Valentina Petkova and Jean-Bernard Zuber for interesting discussions. This work was supported by the ESF grant INSTANS and by the MIUR grant 2007JHLPEZ.

\section*{Note added}
Following the appearance of this article as a preprint, G.~Takacs informed us that complete results for the structure constants of the $D$-series minimal models were obtained in \cite{Runkel}. We checked that, once the different conventions and normalizations are suitably taken into account, the structure constants of \cite{Runkel} for the $D_4$ case exactly coincide with those that we give in the Appendix.

\section*{Appendix}
\label{sec.stru}

The structure constants for the $D$-series unitary minimal models were studied in e.g.\ \cite{scalarconst, ocneanu}. They can be written as 
\beq
C_{ijk}=M_{ijk} \sqrt{D_{\Delta_i\Delta_j\Delta_k}D_{\bar{\Delta}_i\bar{\Delta}_j\bar{\Delta}_k}}\ , 
\eeq
where $\Delta_i$, $\bar{\Delta}_i$  are the left and right conformal weights of $\phi_i$, $D_{\Delta_i\Delta_j\Delta_k}$ and $D_{\bar{\Delta}_i \bar{\Delta}_j \bar{\Delta}_k}$ are structure constants of the corresponding A-series minimal model (for the $D_4$ model this is the $A_5$ model), and $M_{ijk}$ are called relative structure constants.  The constants $D_{\Delta_i \Delta_j \Delta_k}$ and $D_{\bar{\Delta}_i \bar{\Delta}_j \bar{\Delta}_k}$ can be calculated using the formulae given in \cite{Dotsenko}. For the $D_4$ case the relative structure constants $M_{ijk}$ are real numbers on which the $S_3$ symmetry and
\EQ
\sigma_1^\dagger=\sigma_2,\qquad \Omega_1^\dagger=\Omega_2
\label{dagger}
\EN
 impose some restrictions.
The absolute values of these constants were found to be either $0$ or $1$ or $\sqrt{2}$. The latter value occurs in the following cases:
\begin{eqnarray}
|M_{\sigma_2\sigma_1\sigma_1}|=|M_{\sigma_1\sigma_2\sigma_2}|= \nonumber\\
|M_{\Omega_2\sigma_1\sigma_1}|=|M_{\Omega_1\sigma_2\sigma_2}|=
|M_{\sigma_2\sigma_1\Omega_1}|=|M_{\sigma_1\sigma_2\Omega_2}|=\sqrt{2}.
\end{eqnarray}
The signs of the relative structure constants were not completely fixed in \cite{scalarconst, ocneanu}.

It is useful to introduce the combinations 
\begin{eqnarray}
\sigma_\pm & = & \frac{\sigma_1\pm \sigma_2}{\sqrt{2}}\\
\Omega_\pm & = & \frac{\Omega_1\pm \Omega_2}{\sqrt{2}}
\end{eqnarray}
which are even and odd, respectively, under $\ZZ_2$ transformations.
The model has an even part consisting of the modules $I$, $\epsilon$, $X$, $Y$, $\sigma_+$, $\Omega_+$, and an odd part consisting of the modules $\sigma_-$, $\Omega_-$, $J$, $\bar{J}$, $W$, $\bar{W}$. 
The fields in the even part form a closed subalgebra, which is also contained by the $A_5$ model. This means that the even part of the $D_4$ model coincides with a sector of the $A_5$ model.
The $\sqrt{2}$ mentioned above is eliminated from the relative structure constants by using the even and odd combinations introduced above.
 
Since in this paper we study the $D_4$ model with the $\epsilon$ and $\sigma_+$ perturbations, the relative structure constants that we actually need to know are $M_{i\epsilon j}$ and $M_{i \sigma_+ j}$. $C_{ijk}$ with $j=\epsilon$ or $\sigma_+$ is nonzero if both $D_{\Delta_i \Delta_j \Delta_k}$ and  $D_{\bar{\Delta}_i \bar{\Delta}_j \bar{\Delta}_k}$ are nonzero, and the $S_3$ symmetry and (\ref{dagger}) also allow a nonzero value. Then, the nonzero $M_{i\epsilon j}$ are  
\beq
M_{I\epsilon \epsilon}, 
M_{\epsilon\epsilon I},
M_{X\epsilon \epsilon},
M_{\epsilon\epsilon X},
M_{X\epsilon Y},
M_{Y\epsilon X},
M_{\sigma_+\epsilon \sigma_+},
M_{\sigma_+\epsilon \Omega_+},
M_{\Omega_+\epsilon \sigma_+},
\label{c1}
\eeq
for $i$ and $j$ in the even sector, and 
\beq
M_{\sigma_-\epsilon \sigma_-},
M_{\sigma_-\epsilon \Omega_-},
M_{\Omega_-\epsilon \sigma_-},
M_{J\epsilon \bar{J}},
M_{\bar{J}\epsilon J},
M_{J\epsilon W},
M_{W\epsilon J},
M_{\bar{J}\epsilon \bar{W}},
M_{\bar{W}\epsilon \bar{J}},
\label{c2}
\eeq
for $i$ and $j$ in the odd sector. The nonzero $M_{i\sigma_+ j}$ are 
\begin{eqnarray}
M_{I \sigma_+ \sigma_+},
M_{\epsilon \sigma_+ \sigma_+},
M_{\epsilon \sigma_+ \Omega_+},
M_{X \sigma_+ \sigma_+},
M_{X \sigma_+ \Omega_+},\nonumber\\
M_{Y \sigma_+ \sigma_+},
M_{\sigma_+ \sigma_+ I},
M_{\sigma_+ \sigma_+ \epsilon},
M_{\Omega_+ \sigma_+ \epsilon},
M_{\sigma_+ \sigma_+ X},\nonumber\\
M_{\Omega_+ \sigma_+ X},
M_{\sigma_+ \sigma_+ Y},
M_{\sigma_+ \sigma_+ \sigma_+},
M_{\Omega_+ \sigma_+ \sigma_+},
M_{\sigma_+ \sigma_+ \Omega_+},
\label{c3}
\end{eqnarray}
for $i$ and $j$ in the even sector, and
\begin{eqnarray}
M_{\bar{J}\sigma_+\sigma_-},
M_{\bar{J}\sigma_+\Omega_-},
M_{J\sigma_+\sigma_-},
M_{J\sigma_+\Omega_-},\nonumber\\
M_{W\sigma_+\sigma_-},
M_{\bar{W}\sigma_+\sigma_-},
M_{\sigma_-\sigma_+\bar{J}},
M_{\Omega_-\sigma_+\bar{J}},
M_{\sigma_-\sigma_+ J},\nonumber\\
M_{\Omega_-\sigma_+ J},
M_{\sigma_-\sigma_+ W},
M_{\sigma_-\sigma_+ \bar{W}},
M_{\sigma_-\sigma_+\Omega_-},
M_{\Omega_-\sigma_+\sigma_-},
\label{c4}
\end{eqnarray}
for $i$ and $j$ in the odd sector.

The signs of the constants listed in (\ref{c1}), (\ref{c2}), (\ref{c3}) and (\ref{c4}) are partially determined by the $S_3$ symmetry and by (\ref{dagger}). The constants (\ref{c1}) and (\ref{c3}) are equal to $1$ because the even sector coincides with a sector of the $A_5$ model.  

In order to find the signs of the constants (\ref{c2}) and (\ref{c4}) for the odd sector we used the empirical criterion that physically meaningful spectra should be obtained if the $D_4$ conformal theory is perturbed by the  $\epsilon$ and $\sigma_+$ fields. 
The TCSA calculations allow one to fix the signs of all constants (\ref{c2}) and (\ref{c4}) up to a trivial freedom that corresponds to sign changes of the basis vectors. Within one of these equivalent choices the result that we obtained is that the constants (\ref{c2}) are equal to $1$; the constants (\ref{c4}) are equal to $-1$, with the following exceptions:
\beq
\label{exc}
M_{\Omega_-\sigma_+ \bar{J}}=M_{\Omega_-\sigma_+ J}
=M_{\bar{J}\sigma_+ \Omega_-}=M_{J\sigma_+ \Omega_-}=1.
\eeq

\end{document}